\documentstyle[multicol,prb,aps,psfig,preprint]{revtex}
\begin{document}

\title{Modeling the series of ($n\times$2) Si-rich reconstructions of $\beta$--SiC(001): a prospective atomic wire?}

\author{S.A.~Shevlin, A.J.~Fisher\footnote{Email andrew.fisher@ucl.ac.uk} and E.~Hernandez$\dagger$}

\address{Department of Physics and Astronomy, University College
London,\\Gower Street, London WC1E 6BT, United Kingdom}
\address{$\dagger$Institut de Ci\`{e}ncia de Materials de Barcelona - CSIC, Campus de la Universitat Aut\`{o}noma de Barcelona, 08193 Bellaterra, Barcelona, Spain}

\maketitle

\begin{abstract}
We perform {\it ab initio} plane wave supercell density functional
calculations on three candidate models of the (3$\times$2) 
reconstruction of the $\beta$--SiC(001) surface. We find that the
two-adlayer asymmetric-dimer model (TAADM) is unambiguously favored for
all reasonable values of Si chemical potential. We then use structures
derived from the TAADM parent to model the silicon lines that are observed
when the (3$\times$2) reconstruction is annealed (the ($n\times$2)
series of reconstructions), using a density-functional-tight-binding
method. We find that as we increase $n$, and so separate the lines, a
structural transition occurs in which the top addimer of the line
flattens. We also find that associated with the separation of the lines
is a large decrease in the HOMO-LUMO gap, and that the HOMO state becomes
quasi-one-dimensional. These properties are qualititatively and
quantitatively different from the electronic properties of the original
(3$\times$2) reconstruction. 
\end{abstract}
 

\section{Introduction} 

Silicon carbide has long been studied because of its technological potential 
for electronic devices \cite{park,choyke}. While most semiconductor compounds 
exhibit only one stable phase at room temperature, SiC possesses many polytypes. 
Of these polytypes, cubic or $\beta$--SiC has demonstrated its potential for use in 
high-temperature, high-frequency and high-power electronic devices. The lattice 
parameter of $\beta$--SiC possesses a lattice mismatch of $\sim$20\% when compared 
with the lattice parameters of silicon or diamond. This has several consequences: 
first that when $\beta$--SiC is grown as a film on a Si substrate, the interface 
between the two materials possesses an associated strain field. Second, an 
unreconstructed Si or C-terminated SiC surface is under compressive or tensile 
stress. This stress is a driving force for significant reconstructions on both 
surfaces.

Depending on the stoichiometry of the $\beta$--SiC(001) surface, a
large variety of such reconstructions has been found. For silicon-rich 
surfaces (containing more silicon than the Si-terminated bulk structure), 
the series of reconstructions (3$\times$2), (5$\times$2), ...$(n\times$2) has 
been observed \cite{soukiassian97b,hara90}, while for less silicon-rich 
surfaces the $c(4\times$2) \cite{douillard98,soukiassian97,pizzagalli98,lu98,shek96} 
or (2$\times$1) reconstructions are observed \cite{powers92,husken98}. 
For C-terminated surfaces, the $c(2\times$2) reconstruction is 
favoured \cite{powers91,long96}.

It has been found that annealing the highly silicon-rich (3$\times$2) surface at 1100 K 
for several minutes leads to the formation of a grating of very long 
($\sim$1000 \AA), very straight lines at the transition from the (3$\times$2) to 
c(4$\times$2) surface \cite{soukiassian97b,soukiassian97c,semond98,douillard98b}, 
with the separation between the lines of magnitude $\sim$6 \AA . Additional heat 
treatment leads to the removal of this grating and the formation of widely 
separated atomic lines \cite{soukiassian97b,semond98,aristov99}, which we  
refer to as the $(n\times$2) series of reconstructions. These complex ($n\times$2) 
reconstructions consist of two subunits: the structural elements corresponding 
to the area between the lines (believed to be the $c(4\times$2) reconstruction 
\cite{soukiassian97c}), and the structural elements corresponding to the 
lines themselves.

In STM images of these lines \cite{soukiassian97c,semond98} the bright 
regions appear to be $\sim$9 \AA\ wide and the separation of these bright features 
along the line appears to be $\sim$ 6 \AA. The surface lattice vector of
the $\beta$--SiC (001) surface is 3.08 \AA. Thus it is
believed that these lines are formed from the same structural
components as the silicon-rich (3$\times$2) reconstruction, of which
there are several models (see figures~1-3): the additional dimer 
row model (ADRM) \cite{yan95,semond96}, the double dimer row model 
(DDRM) \cite{dayan86,hara94} and the two-adlayer asymmetric-dimer model (TAADM) 
\cite{lu99}. The ADRM has an excess Si surface coverage of 1/3 ML (or one addimer 
per surface unit cell) and has an orientation of (2$\times$3) compared 
to our other candidate models which have an orientation we term 
(3$\times$2), the DDRM possesses a coverage of 2/3 ML (or two addimers 
per surface unit cell), and the TAADM  has a coverage of 1 ML (or three 
addimers per surface unit cell). In the interests of completeness we also 
mention the single dimer row model (SDRM) which is a unit cell which is a 
90$^{\circ}$  rotation of the ADRM and which we find  has a total energy 
2.67 eV higher than the ADRM, and so is thermodynamically unfavorable 
compared to the ADRM; and the two-adlayer-asymmetric rotated dimer 
model (TAARDM) which is a unit cell which resembles the TAADM, but with 
the top addimer rotated so that it is parallel to the second level ad-dimers, 
and which we find has a total energy 0.68 eV higher than the TAADM and so 
is thermodynamically unfavorable when compared with the TAADM. We will not 
consider the SDRM or the TAARDM further.

In Section II we discuss the various calculational methods used in our 
simulations, and how we can compare the grand canonical potentials of the 
various models. In Section III, using {\it ab initio } techniques, we present 
the structural results that we obtain for the (3$\times$2) models, 
the thermodynamic stability of each model, the electronic properties 
of and how they compare to experiment. In Section IV, we perform calculations 
on the series of ($n\times$2) reconstructions modeled by tight-binding, 
using the thermodynamically favoured (3$\times$2) model as the parent, and 
present the details of a change in the structure of the line when $n\geq$7. 
We then present our conclusions in Section V.   

\section{Method}

\subsection{First-principles total energy calculations}

We performed the calculations using {\it ab initio} density functional
theory using the projector augmented wave method (PAW) to handle the
atomic cores \cite{blochl94}. We use the Perdew-Zunger parameterisation 
\cite{perdew81} of the Ceperley-Alder \cite{ceperley80} treatment of the 
uniform electron gas.  We solve the Kohn-Sham equations using the 
Car-Parrinello algorithm \cite{car85}. The main idea of the PAW method 
is to split the all-electron (AE) wavefunction into three parts;
\begin{equation}
|\Psi\rangle = |\tilde\Psi\rangle + \sum_i(|\phi_i\rangle - |\tilde\phi_i\rangle)\langle \tilde p_i|\tilde\Psi\rangle.
\end{equation}
$|\Psi\rangle$ is the AE wave function and $|\tilde\Psi\rangle$ is
a pseudo (PS) wavefunction analogous to the wave functions of the
pseudopotential method, or the envelope functions of the linear
method. The $|\phi_i\rangle$ are a set of AE partial waves within the
core region, while the $|\tilde\phi_i\rangle$ are a set of smooth partial waves
which coincide with the corresponding AE partial
waves outside the core region. $|\tilde p_i\rangle$ are projector
functions localised within the core region which obey the relation
$\langle \tilde p_i|\tilde\phi_j\rangle$ = $\delta_{ij}$.

As there are several models for the (3$\times$2) surface unit cell we
modelled all of them, using the surface slab technique with periodic
boundary conditions. The simulation cells were six and ten atomic
layers deep (six layers were used for preliminary checks, ten layers for 
results quoted here); the bottom two layers were held in their bulk 
configuration while the other layers were allowed to relax, with the 
bottom carbon layer passivated with hydrogen atoms. In our calculations we 
sampled four points of the Brillouin zone, corresponding to the $\Gamma$--point, 
the $J'$--point (in the [$\overline{1}$10] direction in real space), the 
$J$--point (in the [110] direction in real space) and the $K$--point 
(see figure~4). For our structural calculations we used a plane-wave 
cutoff of 20 Rydbergs, but for our electronic spectra and total energy 
calculations we increased the cutoff to 30 Rydbergs. The vacuum 
spacing between slabs was set to 6.08 \AA\ .

The convergence of the simulation with respect to vacuum spacing and  
plane wave cutoff and $k$-point sampling was checked. It was found that 
increasing the vacuum spacing did not affect the resulting relaxed structure. 
The difference between a relaxed cell at a cutoff of 20 Rydbergs and 30 Rydbergs 
was negligible, with minor differences in surface dimer length ($\sim$0.10 \AA\ 
at most) and almost identical ad-dimer length ($\sim$ 0.01 \AA\  difference). 
Dimer buckling was found to be slightly increased for the higher cutoff 
compared to the case with the lower cutoff ($\sim$ 0.06 \AA\ increase). We can 
thus say that our calculations are converged with respect to vacuum spacing and 
plane-wave cutoff.

\subsection{Tight Binding Calculations}

Tight binding methods (TB)~\cite{goringe97} have the advantage of being
computationally much less demanding than first-principles methods, while
still affording relatively high accuracy. The disadvantage originates in
the fact that the approximations involved are usually of a less controllable
nature, and hence are more difficult to improve. Nevertheless, the extreme
simplicity and sometimes surprising accuracy of TB methods make them
a frequently used tool in computational condensed matter and materials
science studies.

We have made use of the TB model known as {\em Density Functional
Tight Binding\/} (DFTB), due to Porezag and coworkers~\cite{porezag95}.
This method goes beyond conventional TB schemes in several ways.
First, an atomic-like basis set is explicitly employed. Because of
this, the model that results is non-orthogonal. In empirical
TB methods it is customary to work in terms of an underlying basis set
(normally assumed to be orthogonal), but only the matrix elements of
the Hamiltonian are used, and the basis set is never explicitly 
constructed. Secondly, the model is constructed by directly evaluating
the Hamiltonian matrix elements within the framework of
Density Functional Theory, albeit using a two-center approximation
(i.e. neglecting environment effects), while in empirical TB models
the matrix elements are adjusted either to empirical or theoretical
data. For a detailed description of the model construction 
procedure, the reader should consult Ref.~\cite{porezag95}.

In DFTB, the total energy of the system under study is calculated as follows. 
The relative positions of the atoms in the system determine the values
of the Hamiltonian and overlap matrix elements. From these, the
Schr\"{o}dinger
equation for the electronic problem can be obtained in matrix form, and
solved by the usual techniques (for example, by matrix diagonalization):
\begin{equation}
\sum_{j\beta} C_{j\beta}^{(n)} (H_{i\alpha,j\beta} - E_n S_{i\alpha,j\beta} )
 = 0.
\label{eq:schroedinger}
\end{equation}
Here indices {\em i\/} and {\em j\/} label atoms, Greek indices label
basis set functions, and {\em n\/} labels eigen-states and eigen-values.
Once the eigen-value problem~(\ref{eq:schroedinger}) is solved, the
{\em band-structure energy\/} is calculated:
\begin{equation}
E_{bs} = 2 \sum_n^{occ} E_n,
\end{equation}
where the sum extends over the occupied eigen-states, and the factor of two
accounts for the degeneracy of spin. To complete the total energy of the
system a repulsive potential contribution is added,
\begin{equation}
E_{rep} = \sum_{j\neq i} V(\mid {\bf R}_j - {\bf R}_i\mid),
\end{equation}
where ${\bf R}_i$ is the position of atom {\em i\/}. This potential 
accounts for the core-core repulsion and the double counting of the
electron-electron energy implicit in the band-structure term 
(see Ref.~\cite{goringe97}).

We have used a DFTB parametrisation consisting of 
four basis functions ($S, P_x, P_y, P_z$) per atom, corresponding to the
four valence orbitals of carbon and silicon. Similar models have been
used with great success by Porezag and coworkers~\cite{porezag95}
to study carbon clusters and hydrocarbons, as well as bulk crystalline 
phases of carbon, amorphous carbon~\cite{kohler95}, Si and SiH 
clusters~\cite{frauenheim95}, crystalline BN~\cite{widany96}, and
SiC structures~\cite{gutierrez96}.

\subsection{Thermodynamical considerations}

We can use the total energies calculated from {\it ab initio} calculations
to discover which of the four candidate models possesses the lowest grand 
canonical potential. We assume that the formation of the different structures 
is determined by thermodynamic factors but we cannot compare the total energies 
of different models with different stoichiometries. A common method of comparing 
structures is to find the chemical potential $\mu_{\rm Si}$ for the silicon
adatoms and then compare the grand potentials \cite{qian88,northup93} 
\begin{equation}
\label{one}
\Omega=E-N_{\rm Si}\mu_{\rm Si}.
\end{equation}
However, because of the experimental conditions that produce the 
(3$\times$2) surface  and because SiC is a two component system, it is nontrivial 
to calculate $\mu_{\rm Si}$. Nevertheless we can estimate a value for the range 
of chemical potential that the silicon ad-atoms experience. We define 
$\Delta E_{\rm Coh}^{\rm Si}$ in terms of $\mu_{\rm Si}$ by the relation
\begin{equation}
\mu_{\rm Si} = E^{\rm atom}_{\rm Si} + \Delta E_{\rm Coh}^{\rm Si},
\end{equation}
where $E^{\rm atom}_{\rm Si}$ is the energy of an isolated silicon
atom (-3.80769 Hartrees in the PAW formalism); $\Delta E_{\rm Coh}^{\rm Si}$
is the corresponding value for the cohesive energy per silicon
atom in our system.

We must decide what physical/chemical process determines 
the value of $\mu_{\rm Si}$ (and hence $\Delta E_{\rm Coh}^{\rm Si}$) in our 
system. We can make several choices. We choose to consider the 
incorporation of Si atoms into bulk SiC. $\Delta E_{\rm Coh}^{\rm Si}$ would 
then become the cohesive energy for a Si atom in SiC. However this 
cannot be calculated directly, because it is impossible to insert 
a single Si into SiC without creating a structural defect of some kind. 
We can, though, estimate the maximum range of $\mu_{\rm Si}$ that the Si 
ad-atoms experience \cite{zyweitz99}. We can write the chemical potential 
per unit cell of an ideal bulk system of SiC as
\begin{equation}
\mu^{\rm bulk}_{\rm SiC}=\mu_{\rm Si}+\mu_{\rm C},
\end{equation}
where $\mu_{\rm Si}$ and $\mu_{\rm C}$ are the (unknown) contributions 
from the Si and C atoms. In extremely Si-rich conditions, $\mu_{\rm Si}$ approaches the 
value of bulk Si. We can consider deviations from this bulk value 
$\Delta\mu_{\rm Si}$=$\mu_{\rm Si}-\mu^{\rm bulk}_{\rm Si}$ (with a 
corresponding definition for $\Delta\mu_{\rm C}$). The allowed 
range is determined by the heat of formation of the SiC compound
\begin{equation}
\Delta H_f =\mu^{bulk}_{\rm SiC}-\mu^{bulk}_{\rm C}-\mu^{bulk}_{\rm Si} \\ =\left(\mu_{\rm Si}-\mu^{bulk}_{\rm Si}\right)+\left(\mu_{\rm C}-\mu^{bulk}_{\rm C}\right)=\Delta\mu_{\rm Si}+\Delta\mu_{\rm C}
\end{equation}
The experimental value of $\Delta H_f$ is 0.72 eV \cite{kubaschewski79}. In 
a Si-rich environment $\Delta\mu_{\rm Si}=0$, and therefore 
$\Delta\mu_{\rm C}=\Delta H_f$. Similarly, in a C-rich environment 
$\Delta\mu_{\rm C}=0$ and $\Delta\mu_{\rm Si}=\Delta H_f$. This 
means that furthest deviation the chemical potential of Si can make 
from the value for bulk Si is 0.72 eV. We show our results as a function of 
$\Delta E^{\rm Si}_{\rm Coh}$ over the range from its value for bulk Si 
$\Delta E^{\rm Si}_{\rm Coh} ({\rm Si})$ to the value obtained by assuming 
that the cohesive energy of SiC can be equally divided between the carbon and 
silicon atoms, we call this value $\Delta E^{\rm Si}_{\rm Coh} ({\rm C})$.

\section{3$\times$2 unit cell}

\subsection{Structural}

It was found from structural calculations that all models we considered 
were locally stable, with silicon adatoms forming addimers. All 
addimers were found to be asymmetric (apart from the case of the TAADM, 
where the second layer addimers were found to be flat). The structure 
obtained for the ADRM is in agreement with previous {\it ab initio} 
\cite{yan95,lu99,pizzagalli99}, and tight-binding work \cite{gutierrez99}, 
with a strongly bound and asymmetric addimer. The structure obtained for 
the TAADM is in agreement with previous work \cite{lu99}. There is an 
alternating arrangement of first layer and second layer addimers as we 
look along [$\overline{1}$10], with the top layer addimer strongly bound and 
asymmetric, and the second level addimers strongly bound and flat. However, 
the structure obtained for the DDRM is in disagreement with both previous 
{\it ab initio} \cite{pizzagalli99}, and tight-binding \cite{gutierrez99} results, 
as we find that both addimers are buckled and quite strongly bound. Our result 
corresponds to one of the DDRM models studied previously \cite{lu99} (the 
LAFM-DDRM structure). The other DDRM models were all found to be of higher total 
energy than this structure, in contradiction with previous work \cite{lu99}. The 
Si atoms of the first Si layer of the SiC proper were also found to be dimerised 
in all unit cells, in agreement with previous work 
\cite{lu99,pizzagalli99,gutierrez99,kitabatake96,kitabatake98}. Table~I provides 
further information about the equilibrium structure of each model.

\subsection{Thermodynamics}

We use calculated {\it ab initio} total energies, to find $\Omega$ from equation 
$(\ref{one})$ between the various models. The results presented (see figure~5) 
are for our most converged simulations (30 Rydberg cutoff, 4 $k$-points sampled). 
We first of all start off with the assumption that the Si adatoms 
are in equilibrium with the bulk. As can be seen, all crossover points (that 
is the values of $\Delta E_{\rm Coh}^{\rm Si}$ for which $\Omega$ is equal 
for two separate models) are outside the maximum allowable range of 
$\Delta E_{\rm Coh}^{\rm Si}$ as determined by $\Delta H_f$. We can thus 
unambiguously apply our calculations of $\Omega$ to determine which 
model is thermodynamically preferred. Over the entire range of allowable 
chemical potentials, the TAADM model is favored.

Our results are in agreement with the previous work of Lu {\it et al.} 
\cite{lu99}, where it is found that the TAADM is the thermodynamically preferred 
model throughout the entire allowable range of 
$\Delta E_{\rm Coh}^{\rm Si}({\rm SiC})$. Regarding the relative stability of the 
ADRM and DDRM , we agree with other work \cite{pizzagalli99,gutierrez99}, where it 
is suggested that the ADRM is the favoured model, although the TAADM is not considered. 
We find that the ADRM is only favored for $\Delta E_{\rm Coh}^{\rm Si}$ which are 
close to $\Delta E_{\rm Coh}^{\rm Si}({\rm SiC})$. But there is no 
evidence for a (2$\times$3) reconstruction which is less silicon-rich than 
the $c(4\times$2) reconstruction. As we find that the c(4$\times$2) 
reconstruction is only preferred over a narrow range of $\mu$ when the 
ADRM is included, (contrary to observations), we conclude that the 
ADRM does not occur in practice, as it would prevent the formation of the 
c(4$\times$2) reconstruction. We can therefore rule out the ADRM.

We have therefore found from our {\it ab initio} calculations of the grand  
potential that the TAADM is preferred over a large range of chemical potential 
and the other two models can be discounted, with the DDRM found to be never 
preferred and the ADRM only valid for a range of chemical potential which is 
non-physical.

\subsection{Electronic}

We can calculate the differences (the `dispersion') between the Kohn-Sham 
eigenvalues calculated at different points of the Brillouin zone, and 
compare these with recent photoemission experiments performed on the 
(3$\times$2) surface \cite{bermudez95,lubbe98,yeom97,yeom98}. Our results, 
and equivalent results in the literature are presented in Table~2. The magnitude 
of dispersion in the majority of cases is approximately equal to the magnitude 
of dispersion measured \cite{yeom98} ($\sim$0.2eV). The exception is for the 
DDRM, where we find a large dispersion along $\Gamma-J$. We attribute this to our 
modeling of a perfectly ordered periodic (3$\times$2) surface, whereas the 
(3$\times$2) surface has been observed to be almost perfectly disordered 
\cite{hara96,hara99}, in the sense that there is no observable correlation 
between the addimer tilt in one (3$\times$2) unit cell, and the next (in either 
the $\times$3 or $\times$2 direction). The measured dispersion of the surface 
state bands is $\sim$0.2 eV along $\Gamma-J'$, with no measured dispersion 
along $J$ \cite{yeom98}. We find that the only model which reflects this anisotropy 
of dispersion is the ADRM, with the DDRM and the TAADM both possessing larger 
dispersions along $\Gamma-J$ and smaller dispersions along $\Gamma-J'$, contrary 
to experiment. As seen in Table~II, there are some disagreements among the different 
calculations. We especially highlight the discrepancy between our results and those of 
Lu {\it et al.} \cite{lu99} for the energetically preferred TAADM model. 
However we find that the HOMO state for the TAADM exists mostly in the vacuum of the 
simulation slab, and suggest that as we use plane waves to describe our electron 
wavefunction, as opposed to the Gaussian orbitals used in other work \cite{lu99}, 
our dispersion results more accurately describes the dispersion of the surface state.

\section{Larger unit cells ($n\times$2)}

We can model the ($n\times$2) series of reconstructions by the mixing of 
two sets of reconstructions, the (3$\times$2) and the $c(4\times$2). 
We use the thermodynamically favoured TAADM model as the parent reconstruction 
of the lines themselves. We use the favoured MRAD model \cite{lu98,shevlin00} as 
the surface between the lines \cite{lu00}. Due to the large size of the 
reconstructions, {\it ab initio} methods are inefficient for modelling these 
surfaces. We therefore used the non-self-consistent density-functional 
tight-binding method to simulate these surfaces \cite{porezag95,frauenheim95}. 
We performed simulations for $n$=3,5,7,9 and 11 sampling both the $\Gamma$ 
point and a (221) $k$--point mesh generated by the Monkhorst-Pack scheme 
\cite{monkhorst76}. It was found that structural calculations are converged 
for this choice of $k$--point sampling scheme. All simulation cells are ten 
atomic layers thick, with the bottom carbon layer terminated with hydrogen to 
avoid artificial charge transfer. The results can be neatly summarised in Figure~6.

As can be seen, a comparison of {\it ab initio} and tight-binding simulations 
for the (3$\times$2) surface unit cell shows that although the addimer bond 
length $a_1$ is different in the two cases, with the {\it ab initio} addimer 
length found to be 2.31\AA\ and the tight-binding addimer length found to be 
2.62\AA\ , we find a similar buckling $\Delta z$, (0.58\AA\ and 0.53\AA\ 
respectively). We find that as we increase the width of the surface unit 
cell from $n$=3, to $n$=5, the asymmetry and length of the weak addimer 
$a_1$ remain approximately the same (this is qualitatively in agreement with 
previous theoretical work \cite{lu00}, wherein it is found that the asymmetry 
and length of the top ad-dimer remains the same). However, we find that for 
$n\geq$7, the top addimer becomes symmetric and the addimer bond length becomes 
shorter with a bond length of 2.35\AA\ . This matches STM topographs of the lines, 
which show that the lines are composed of symmetric units \cite{douillard98b,derycke}. 
The flat addimer of the (7$\times$2) unit cell has a shorter bond length than the 
buckled addimer of the (3$\times$2) or (5$\times$2) because there is a stronger 
$\sigma$ bond and a much stronger $\pi$ bond. The bond angles between the top addimer 
and the top-adlayer-to-second-adlayer bonds all become $\sim 109.5^\circ$, compared 
to $\sim 85^\circ$ and $125^\circ$ for the buckled addimer case. It was also found 
that if the addimer is buckled and then left to relax in a unit cell where all other 
atoms are already in their relaxed positions, then the addimer remains buckled, albeit 
with a higher energy than the flat addimer. That is, the structural transition may 
be kinematically limited. The buckled addimer structure is 0.25 eV higher in 
total energy than the flat addimer structure; however we point out that this is less 
than the average thermal energy of all the surface silicon atoms at room temperature. 
The flat and buckled addimer structures could both be accessible to the surface.

Analysis of the bond order (off-diagonal density matrix elements $\rho^i_j$) between 
the top adatom and adatoms of the second adlayer show that as the addimer becomes 
flat this bond becomes much stronger, with larger $\sigma$ and $\pi$ components. 
Associated with this stronger bond between top adlayer and second adlayer is 
the loss of an electron from the top addimer. This electron is transferred to the surface 
silicon atoms directly below the addimer as shown in Figure~6. 

We also find that coincident with the isolation of the separate lines, there is a 
change in the HOMO-LUMO (highest-occupied-molecular-orbital lowest-unoccupied-molecular-
orbital) gap. Analysis of the electronic structure of the unit cell was performed 
using a (551) $k$--point Monkhorst-Pack net. Within our tight-binding formalism, we 
find that the HOMO-LUMO gap of the (3$\times$2) unit cell varies from 1.48 eV at the 
zone centre, to 1.14 eV at the zone boundary. For the (7$\times$2) unit cell, we find 
that the HOMO-LUMO gap is decreased by a considerable amount, from 0.87 eV at the 
zone centre to 0.30 eV at the zone boundary. We also find that the anisotropy of 
dispersion of the HOMO state changes; for the (3$\times$2) unit cell the HOMO 
state disperses by 0.05 eV along $\Gamma-J'$ and by 0.43 eV along $\Gamma-J$, 
while for the (7$\times$2) unit cell we find that the dispersion of 
the HOMO state is 0.43 eV along $\Gamma-J'$ and 0.05 eV along $\Gamma-J$. As 
the HOMO state is associated with the top addimer, and as this is now more strongly 
bound to the second level addimers, this means that there there are now strong 
connections along the line. Therefore the HOMO state of the isolated lines shows strong 
quasi-one-dimensional behaviour when compared with the interacting silicon lines that 
constitute the (3$\times$2) and (5$\times$2) reconstructions. 

We observe three changes that happen together: that the HOMO state becomes
quasi-one-dimensional in character, that the top adatom of the buckled
addimer moves down, and that there is a transfer of an electron from the top
adlayer to the top layer of the silicon carbide crystal proper. We offer one possible
rationalisation for this structural transition: that as the lines
become separated, the HOMO state becomes quasi-one-dimensional. This makes the
HOMO state interact more strongly along the line, which means that there 
is an extra contribution to the bonds to the addimers of the second
layer. This forces the top adatom closer to the surface, and flattens the buckle
of the top addimer. As the dangling bond state rises in energy as the addimer 
becomes flat, a state of the top adatom becomes depopulated, and an electron
leaves the top addimer, to reside in the silicon atoms of the top layer of the
silicon carbide crystal directly below the addimer. However we are aware that
the charge transfer which seems to be involved in this structural transition
could be inaccurately treated, as the tight-binding method we use is a 
non-charge-self-consistent scheme.

\section{Conclusions} 

We have performed detailed electronic structure calculations on several 
different models of the silicon-rich (3$\times$2) reconstruction of cubic 
silicon carbide. It was found that the 1 ML TAADM model was the preferred 
model for the (3$\times$2) reconstruction over the entire range of allowable 
chemical potential. Our calculated dispersion values for the TAADM contradict 
experiment however, we find that the only model that matches the 
photoemission data \cite{yeom98} is the ADRM. The mapping between observed 
surface states and one electron Kohn-Sham eigenvalues is not well defined 
however. We conclude that the TAADM is the theoretically preferred model.

We have also used our DFT electronic structure calculations as the basis 
for a tight-binding analysis of the ($n\times$2) series of reconstructions, 
that is those reconstructions that correspond to the silicon lines 
observed on the $c(4\times$2) surface 
\cite{soukiassian97b,soukiassian97c,semond98,douillard98b}. We find that there 
is a structural transition associated with a critical value of $n$, where if $n\geq$7, 
then the top addimer of the TAADM becomes flat. We find that the HOMO-LUMO gap 
associated with the increasing separation of the lines is reduced, so that within 
the tight-binding formalism the ($n\geq7\times$2) reconstructions are 
narrow bandgap semiconductors. We also find that the HOMO state associated with 
the flat addimer is now confined to the line and is strongly dispersed along 
the line, i.e. it displays quasi-one-dimensional behaviour. Our results thus show 
that the electronic and structural properties of the ($n\times$2) series of 
reconstructions, corresponding to widely separated silicon lines, are very different 
from the electronic and structural properties of the (3$\times$2) surface. 
Detailed experimental data on the physical properties of the ($n\times$2) series 
of reconstructions is needed to verify these theoretical results.

\section*{Acknowledgements}
This work was supported by the UK Engineering and Physical Sciences
Research Council.  We would like to thank Herv\'e Ness, John Harding,
David Bowler, Marshall Stoneham and Tony Harker for a number of helpful discussions.

\begin{table}

\caption{Equilibrium bond lengths for the addimers and buckling of 
addimers (magnitude) in our total energy calculations. All distances are 
in \AA\ . See Figures~1-3 for more.}
\label{Table 1}

\begin{tabular}{cccc}
Model &  Addimer bond length & Buckling of addimer & Surface dimer length\\\hline 
ADRM & 2.30 ($a_1$) & 0.50 & 2.59 ($d_1$), 2.47 ($d_2$), 2.49 ($d_3$)\\ 
DDRM & 2.35 ($a_1$), 2.34 ($a_2$) & 0.51, 0.53 & 2.40 ($d_1$), 2.41 ($d_2$)\\ 
TAADM & 2.31 ($a_1$), 2.44 ($a_2$), 2.42 ($a_3$) & 0.58, 0.01, 0.00 & 2.41 ($d_1$), 2.41 ($d_2$)\\
\end{tabular}

\end{table}

\begin{table}

\caption{`Dispersion' of HOMO state along [$\overline{1}$10] ($\Gamma-J'$) 
and [110] ($\Gamma-J$ directions, for various models. Also shown are values 
found in the literature. All values are in eV.}

\begin{tabular}{cccc}
Author & ADRM & DDRM & TAADM\\\hline
This work & 0.18 ($\Gamma J'$), 0.00 ($\Gamma J$) & 0.01 ($\Gamma J'$), 0.34 ($\Gamma J$) & 0.13 ($\Gamma J'$), 0.21 ($\Gamma J$)\\
Pizzagalli {\it et al} \cite{pizzagalli99} & $\leq$ 0.10 ($\Gamma J'$), $\leq$ 0.10 ($\Gamma J$) & 1.00 ($\Gamma J'$), $\leq$ 0.10 ($\Gamma J$) & \\
Lu {\it et al} \cite{lu99} & 0.20 ($\Gamma J'$), $\sim$ 0.00 ($\Gamma J$) & $\sim$ 0.00 ($\Gamma J'$), $\sim$ 0.50 ($\Gamma J$) & 0.37 ($\Gamma J'$), $\sim$ 0.00 ($\Gamma J$)\\
\end{tabular}

\end{table}

\begin{figure}[htb]\narrowtext
\centering
\vspace*{6cm}
\includegraphics{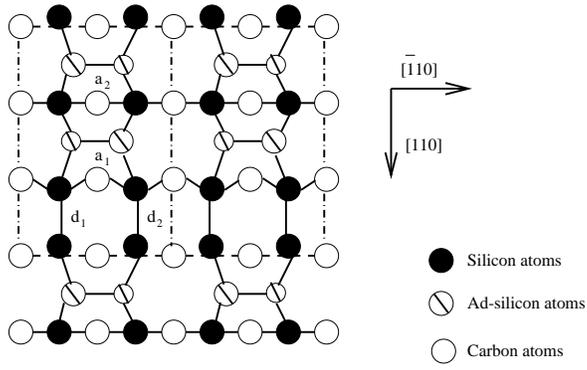}
\caption{Schematic illustration of the DDRM (Anti-ferromagnetic configuration).} 
\label{Figure 1}
\end{figure}

\begin{figure}[htb]\narrowtext
\centering
\vspace*{6cm}
\includegraphics{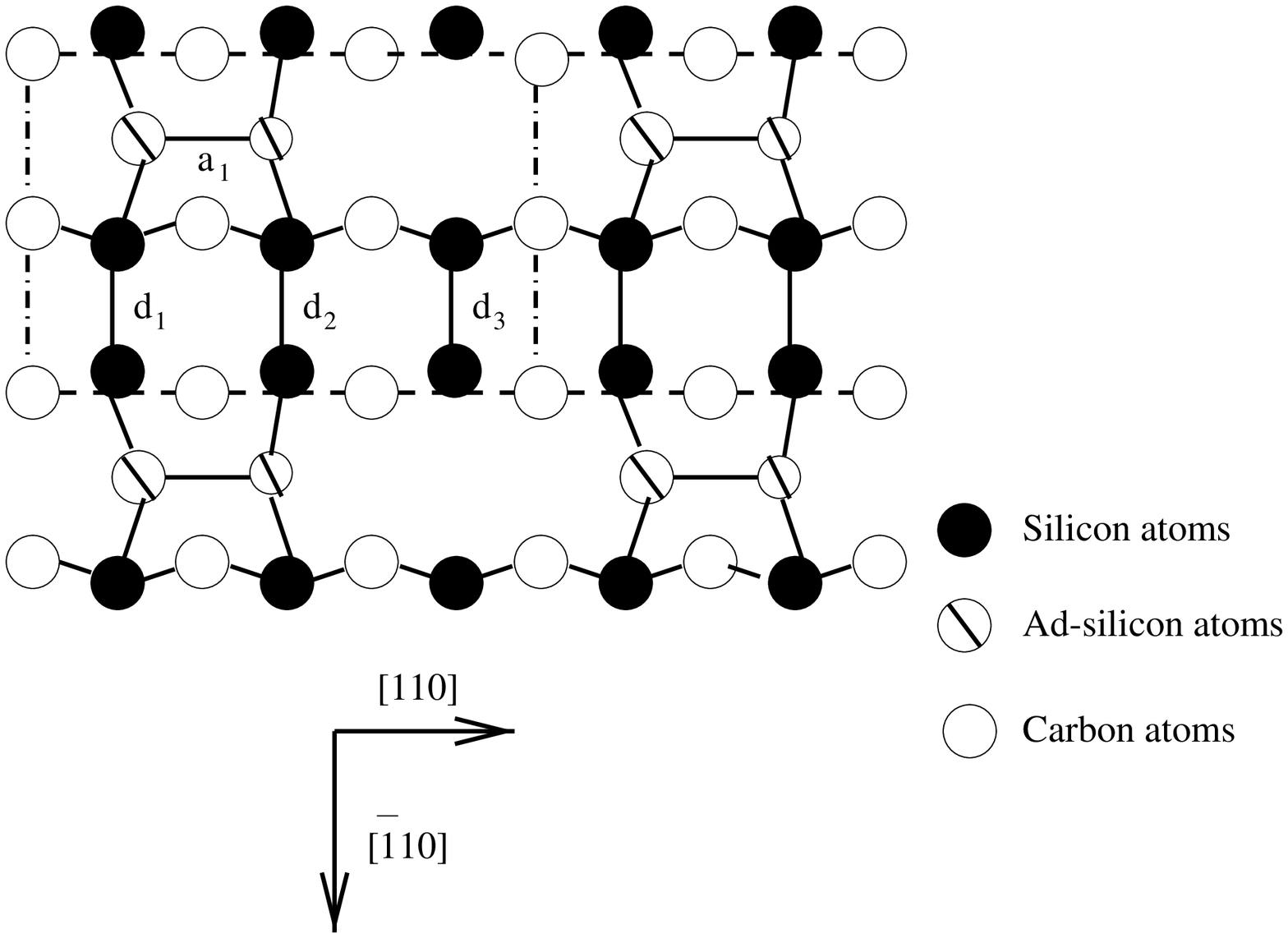}
\caption{Schematic illustration of the ADRM.} 
\label{Figure 2}
\end{figure}

\begin{figure}[htb]\narrowtext
\centering
\vspace*{6cm}
\includegraphics{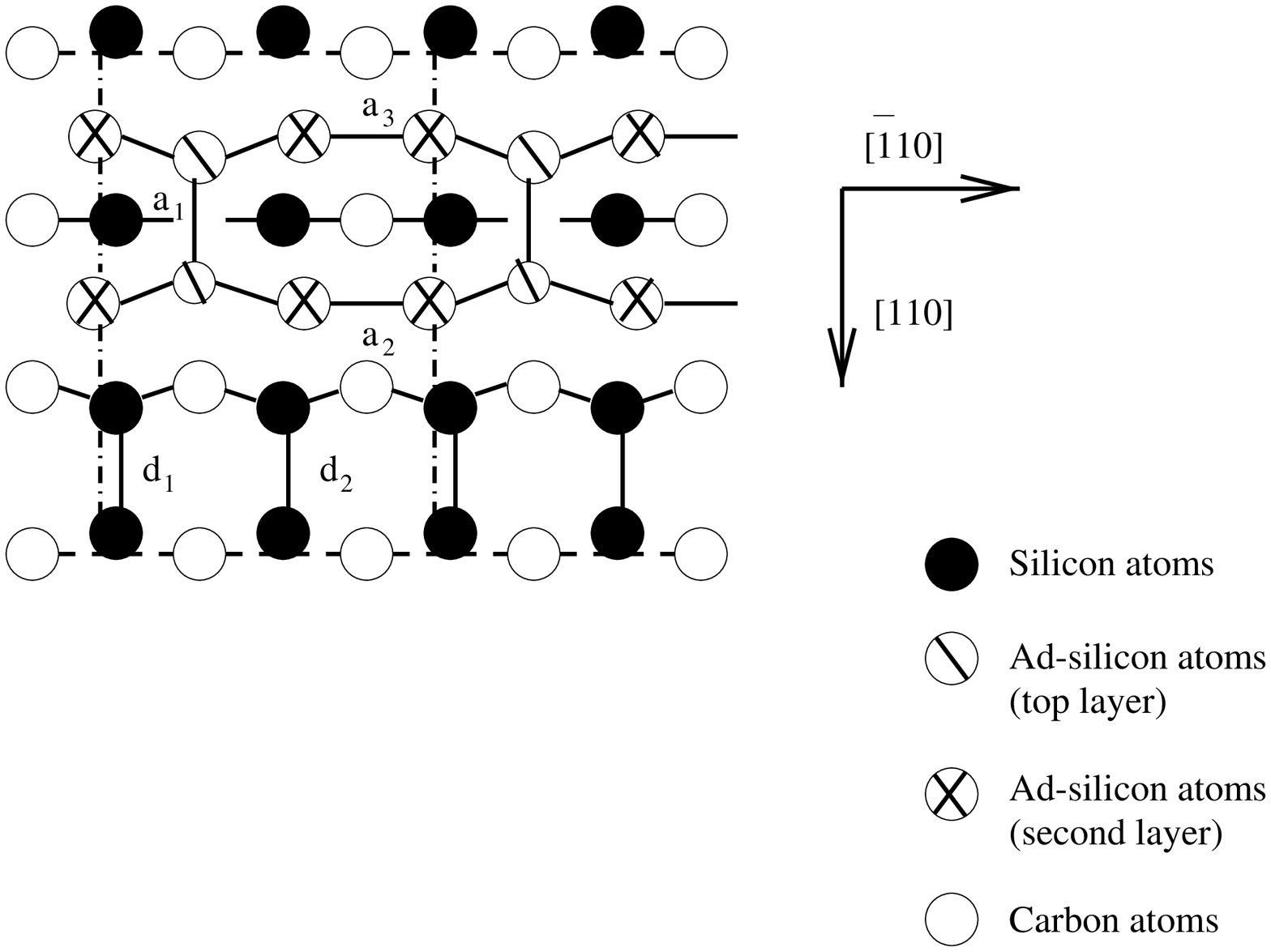}
\caption{Schematic illustration of the TAADM.} 
\label{Figure 3}
\end{figure} 

\begin{figure}[htb]\narrowtext
\centering
\vspace*{7cm}
\includegraphics{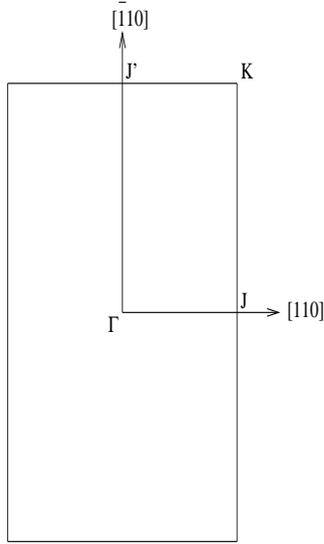}
\caption{Depiction of the surface Brillouin zone and the four special 
$k$-points sampled.} 
\label{Figure 4}
\end{figure}

\begin{figure}[htb]\narrowtext
\centering
\vspace*{7cm}
\includegraphics{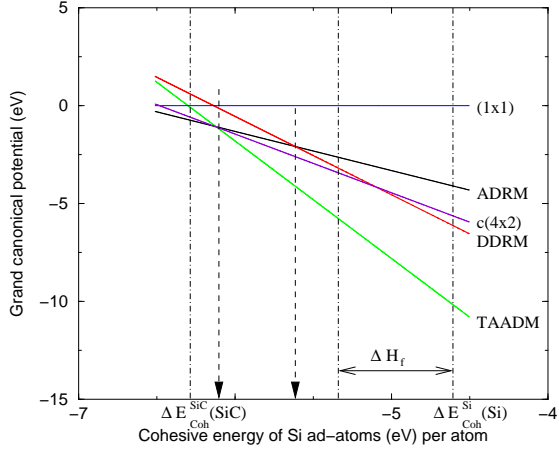}
\caption{Plot comparing the grand canonical potentials $\Omega$ for the three 
candidate models, the ideal (1$\times$1) surface and a (3$\times$2) slice 
of the $c(4\times$2) reconstruction, as a function of cohesive energy per Si atom 
$\Delta E_{\rm Coh}^{\rm Si}$. The zero of the grand canonical potential corresponds 
to the grand canonical potential of the ideal (1$\times$1) surface. Arrows pointing 
downwards indicates the critical value $\Delta E^*_{\rm Coh}$ where $\Omega$ is 
equal for a pair of models. Also indicated are the cohesive energies per atom of 
silicon and silicon carbide and the range of $\Delta H_f$, to illustrate the 
maximum range of allowable cohesive energies, for the case where the Si atoms are 
assumed to be in equilibrium with the bulk.} 
\label{Figure 5}
\end{figure}

\begin{figure}[htb]\narrowtext
\centering
\vspace*{9cm}
\includegraphics{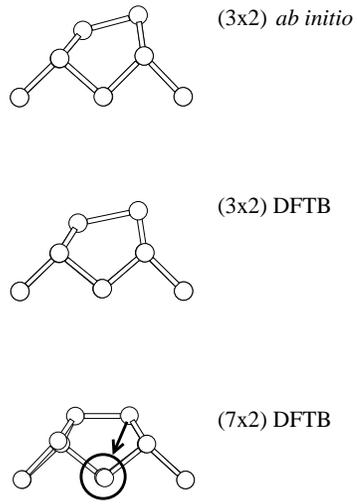}
\caption{Illustration of the different addimer structures we obtain for different 
unit cells and different ,methods of calculation ({\it ab initio} and DFTB). Note 
that the addimer for the (7$\times$2) is flat, compared to the (3$\times$2) unit 
cell (obtained by both methods). The circle on the (7$\times$2) structure indicates 
where the electron from the top addimer localizes when the transition from 
buckled to flat addimer occurs.} 
\label{Figure 6}
\end{figure}

\end{document}